\documentclass{article}
\usepackage{graphicx}
\usepackage{amsfonts}
\usepackage{amsmath}
\usepackage{bm}
\usepackage{epstopdf}
\usepackage{booktabs}
\usepackage{lipsum}
\usepackage{amssymb}
\usepackage{algorithm}
\usepackage{algorithmic}
\usepackage{threeparttable}
\usepackage{multirow}

\begin{document}

\title{Search-free DOA Estimation Method Based on Tensor Decomposition and Polynomial Rooting for Transmit Beamspace MIMO Radar}

\author{Feng Xu, Xiaopeng Yang and Tian Lan}
\maketitle

\begin{abstract}
In order to improve the accuracy and resolution for transmit beamspace multiple-input multiple-output (MIMO) radar, a search-free direction-of-arrival (DOA) estimation method based on tensor decomposition and polynomial rooting is proposed. In the proposed method, a 3-order tensor is firstly designed to model the received signal of MIMO radar on the basis of the multi-linear property. Then, the factor matrix with target DOA information is obtained by the tensor decomposition via alternating least squares (ALS) algorithm, and subsequently the DOA estimation is converted into the independent minimization problem. By exploiting the Vandermonde structure of the transmit steering vector, a polynomial function is constructed to solve the minimization problem via polynomial rooting. The factor matrix contained in the coefficients of the polynomial can be regarded as a block matrix in the generalized sidelobe canceller (GSC), which accordingly forms a unique deep null in the direction of target in the transmit beampattern. The proposed method can obtain the DOA estimation without the requirements of spectrum searching or transmit beamspace matrix design, which is different from the conventional DOA estimation techniques. The effectiveness of the proposed method is verified by the simulations.
\end{abstract}

\section{Introduction}

Multiple-Input Multiple-Output (MIMO) radar radar has been the focus of intensive research for over a decade\cite{1,2,3,4,5}. Direction-of-arrival (DOA) estimation is one of the most fundamental topics among these researches\cite{6,7,16,18,19,20,41}. Much of the literature has generalized classic DOA estimation algorithms from conventional phased array radar to MIMO radar, such as multiple signal classification (MUSIC)\cite{6,7,8}, root-MUSIC\cite{16,19,41}, and estimation of signal parameters via rotational invariance technique (ESPRIT)\cite{9,10,15,18}. Meanwhile, the multi-linear property of the received signal in MIMO radar has been demonstrated\cite{12,13,14,40}. Methods like parallel factors analysis (PARAFAC) \cite{8} can be applied to decompose the factor matrices of a tensor to conduct DOA estimation conveniently. Moreover, it has been shown that the DOA estimation performance of the tensor decomposition-based methods is better than that of the covariance matrix-based methods\cite{8,14}.

In transmit beamspace MIMO radar, with a number of waveforms less than the number of transmit elements, the emitted energy can be focused on a given region\cite{7,11}. This trade-off between the waveform diversity and spatial diversity mitigates the deterioration of targets gain caused by the omnidirectional transmit beampattern. Nevertheless, the use of transmit beamspace technique destroys the Vandermonde structure of the transmit steering matrix. In order to conduct DOA estimation, spectrum searching-based method like MUSIC can be applied at the cost of high computational complexity. The transmit beamspace matrix with special structure has been designed in \cite{11,22,23,25} to enforce rotational invariance property (RIP) in transmit beamspace MIMO, while several subarrays with identical transmit beamspace matrix has been introduced in \cite{21}. Then an ESPRIT-aided DOA estimation is performed. The aforementioned methods rise either the complexity of the transmit beamspace matrix design or other additional computation, which is inappropriate in some scenarios. To solve this problem, a search-free DOA estimation method for transmit beamspace MIMO radar is investigated.

In this paper, a search-free DOA estimation method based on tensor decomposition and polynomial rooting is proposed to improve the accuracy and resolution for transmit beamspace MIMO radar. Specifically, a 3-order tensor is firstly designed to model the received signal of MIMO radar on the basis of the multi-linear property. The factor matrix with target DOA information is obtained by the tensor decomposition via alternating least squares (ALS) algorithm. Then, the DOA estimation is converted into the independent minimization problem. By using the Vandermonde structure of the transmit steering vector, a polynomial function is constructed to estimate the target DOA via polynomial rooting. The proposed method can obtain the DOA estimation without the requirement of spectrum searching or transmit beamspace matrix design. Simulation results show that the method can achieve better accuracy and higher resolution for the transmit beamspace MIMO radar compared with the conventional DOA estimation techniques.

\section{Signal Model}
We firstly present a useful conclusion that will be used later. For any matrices ${\bf{A}} \in {\mathbb{C}^{M \times N}}$, ${\bf{C}} \in {\mathbb{C}^{N \times P}}$, and diagonal matrix ${\bf{B}} = diag({\bf{b}}) \in {\mathbb{C}^{N \times N}}$, where the operator $diag( \cdot )$ returns a square diagonal matrix with diagonal elements equal to its vector argument, we have the following
 \begin{equation}
vec\left\{ {{\bf{ABC}}} \right\} = \left( {{{\bf{C}}^H} \odot {\bf{A}}} \right){\bf{b}}
\end{equation}
where $vec\{\cdot\}$ stacks the elements of a matrix one by one to a column vector, $ \odot $ is the Khatri-Rao product, and ${\left( \cdot \right)^H}$ represents the Hermitian transpose.

Consider a collocated MIMO radar with $M$ transmit elements organized in a uniform linear array (ULA) and $N$ receive elements with arbitrary array geometry within a fixed aperture. The distance between transmit elements is ${d_t}$. The $M \times 1$ transmit steering vector can be denoted by ${\bm{\alpha }}(\theta ) \triangleq {[1,{e^{ - j\frac{{2\pi }}{\lambda }{d_t}\sin \theta }},\cdots,{e^{ - j\frac{{2\pi }}{\lambda }(M - 1){d_t}\sin \theta }}]^T}$, where ${( \cdot )^T}$ denotes the transpose operator and $\theta $ represents the target direction. Similarly, the $N \times 1$ receive steering vector of $N$ receive elements is given as ${\bm{\beta }}(\theta ) \triangleq {[1,{e^{ - j\frac{{2\pi }}{\lambda }{x_2}\sin \theta }},\cdots,{e^{ - j\frac{{2\pi }}{\lambda }{x_N}\sin \theta }}]^T}$, where $\left\{ {\left. x_n \right|{\rm{0}} \le {x_n} \le {D_r},n = 1,\cdots,N} \right\}$ are the coordinates of the receive elements and ${D_r}$ is the aperture of the receive array.

Let ${\bf{S}}(t) \triangleq [{S_1}(t),{S_2}(t),\cdots,{S_M}(t)]^T$ be the $M \times 1$ vector of the pre-designed waveforms. For any two different transmit waveforms in ${\bf{S}}(t)$, the orthogonality property needs to be satisfied, i.e., $\int_T {{\bf{S}}(t){{\bf{S}}^H}(t)dt}  = {{\bf{I}}_M}$, where $T$ denotes the radar pulse duration and ${{\bf{I}}_M}$ is the $M \times M$ identity matrix. The matrix of transmit waveforms is denoted by ${\bf{Z}}(t) = {{\bm{\alpha }}^H}(\theta ){\bf{S}}(t)$. Assuming there are $L$ targets at ${\theta _l}, l = 1,2,\cdots,L$, the received signal of all reflections from the targets can be formulated as ${\bf{y}}(t) = \sum\limits_{l = 1}^L {\sigma _l^2{\bm{\beta }}({\theta _l}){{\bm{\alpha }}^H}({\theta _l}){\bf{S}}(t)}  + {\bf{n}}(t)$, where $\sigma _l^2$ is the radar cross section (RCS) fading coefficient (can be regarded as a function of target RCS) and ${\bf{n}}(t)$ is the zero-mean white Gaussian noise. The output of the matched-filter in matrix form is
\begin{equation}
{\bf{Y}} = {\bf{B\Sigma }}{{\bf{A}}^H} + {\bf{N}} \label{b}
\end{equation}
where ${\bf{B}}\triangleq {[{\bm{\beta }}({\theta _1}),{\bm{\beta }}({\theta _2}),\cdots,{\bm{\beta }}({\theta _L})]_{N \times L}}$, ${\bf{A}} \triangleq {[{\bm{\alpha }}({\theta _1}),{\bm{\alpha }}({\theta _2}),\cdots,{\bm{\alpha }}({\theta _L})]_{M \times L}}$ are the receive and transmit steering matrices, respectively, ${\bf{N}}$ is the noise residue, and ${\bf{\Sigma }} = diag({\bf{c}})$ is a diagonal matrix with ${\bf{c}} \triangleq [\sigma _1^2,\cdots,\sigma _L^2]^T$. The DOA estimation is to determine all $\theta_l$ from the observation of ${\bf Y}$.

\section{Search-free DOA Estimation via Tensor Decomposition and Polynomial Rooting}
In this section, we firstly design a tensor model of the received signal for transmit beamspace MIMO radar, and obtain the factor matrix with target DOA information via tensor decomposition. Then, the DOA estimation is converted into a polynomial rooting problem. The relationship between the constructed polynomial function and MIMO radar transmit beampattern is also illustrated. Hence, a search-free DOA estimation method based on tensor decomposition and polynomial rooting is proposed.

\subsection{Tensor Model of Received Signal and Tensor Decomposition}
Let ${\bf{W}} \triangleq {\left[ {{\bf {w}}_1,{\bf {w}}_2,\cdots,{\bf {w}}_K} \right]}_{M \times K}$ be the transmit beamspace matrix \cite{7,11,22}. The transmitted signals are then ${{\bf{W}}^H}{\bf A}$ for $L$ targets. Substituting ${{\bf{W}}^H}{\bf A}$ into \eqref{b} and considering the Doppler effect, the received signal in $q$-th pulse can be written as
\begin{equation}
{\bf{Y}} = {\bf{B\Sigma }}diag({\bm{\gamma }}_q){({{\bf{W}}^H}{\bf{A}})^H} + {\bf{N}}\label{data}
\end{equation}
where ${\bm{\gamma }}_q \triangleq [{e^{j2\pi {f_1}q{T}}},{e^{j2\pi {f_2}q{T}}},\cdots,{e^{j2\pi {f_L}q{T}}}]^T$ is the vector of Doppler shifts and $f_l$ is the Doppler frequency of $l$-th target. Vectorizing \eqref{data} into a $KN \times 1$ vector to obtain
\begin{equation}
{\bf{y}}_{q} = \left[ {\left( {{{\bf{W}}^H}{\bf{A}}} \right) \odot {\bf{B}}} \right]{{\bf{c}}_{q}} + {\bf{\tilde n}}_q
\end{equation}
where ${\bf{c}}_{q} = {{\bf{c}}} \diamond {\bm{\gamma }}_q$, ${\bf{\tilde n}}_q$ is the corresponding noise residue, and $\diamond$ denotes the Hadamard product. Assuming that $Q$ pulses are utilized in a single coherent processing interval (CPI), the received signal can be concatenated by ${\bf{Y}}_w \triangleq [{{\bf{y}}_{1}},{{\bf{ y}}_{2}},\cdots,{{\bf{y}}_{Q}}]$, or equivalently, by
\begin{equation}
{\bf{Y}}_w = \left[{{{\left( {{{\bf{W}}^H}{\bf{A}}} \right)}} \odot {\bf{B}}} \right]{{\bf{C}}^T} + {\bf{\tilde N}} \label{f}
\end{equation}
where ${{\bf{C}}} \triangleq [{\bf{c}}_{1},{\bf{c}}_{2},\cdots,{\bf{c}}_{Q}]^T$ and ${\bf{\tilde N}} \triangleq [{{\bf{\tilde n}}_1},{{\bf{\tilde n}}_2},\cdots,{{\bf{\tilde n}}_Q}]$.

Note that ${\bf{Y}}_w$ can be regarded as the matrix form of a 3-order tensor ${{\cal Y} \in {{\mathbb{C}}^{{K} \times {N}\times {Q}}}}$ unfolded across the third dimension. Referring to \cite{13,14}, $\cal Y$ is written as ${\cal Y}= \sum\limits_{l = 1}^L {{{\bf{x}}_l} \circ {{\bm{\beta}}_l} \circ {{\bm{\eta}}_l}}+{\cal N} \triangleq  [[ {{\bf{X}},{\bf{B}},{\bf{C}}}
 ]]+ {\cal N}$, where $\circ$ is the outer product, ${\bf{X}} = {{\bf{W}}^H}{\bf{A}}$, $\cal N$ is the noise tensor, and ${\bf{x}}_l, {{\bm{\beta}}_l},{{\bm{\eta}}_l}$ are the $l$-th column of ${\bf{X}},{\bf{B}},{\bf{C}}$, respectively. To decompose all factor matrices of $\cal Y$ simultaneously, the ALS algorithm\cite{13} is applied, i.e., solving the following problem alternatingly among three factor matrices ${\bf{X}},{\bf{B}}$ and ${\bf{C}}$
\begin{equation}
\mathop {\min }\limits_{{\bf{\hat X}}} \left\| {{{{\cal Y}}_{(1)}} - \left[ {({\bf{C}} \odot {\bf{B}}){{{\bf{\hat X}}}^T}} \right]} \right\|_F^2\\
\label{ALS}
\end{equation}
where ${{{{\cal Y}}_{(1)}}}$ denotes matrix form of ${\cal Y}$ across the first dimension, $\| \cdot \|_F$ denoted the Frobenius norm, and $\bf{\hat X}$ is the estimation of $\bf{X}$. During each alternating step, the objective function similar to the one in \eqref{ALS} is quadratic with respect to the optimized matrix parameter ($\bf X$, $\bf B$ or $\bf C$). After tensor decomposition, ${\bf \hat X}$ is the desired factor matrix contains the target DOA information.

\subsection{DOA Estimation via Polynomial Rooting}
In conventional MIMO radar, the transmit beamspace matrix $\bf W$ can be regarded as the identity matrix. Hence, the factor matrix after the tensor decomposition is Vandermonde (${\bf X} = {\bf A}$) and the DOA information can be obtained from it conveniently. In transmit beamspace MIMO radar, to conduct DOA estimation \cite{11,25}, the RIP is enforced by concatenating the transmit beamspace matrix with its flipped-conjugate version, i.e., $[{\bf{W}},{\bf{\bar W}}]$, where ${\bf \bar W} \triangleq [{\bf \bar w}_1,{\bf \bar w}_2 \cdots {\bf \bar w}_K]$, \ ${\bf \bar w}_k(m) = {\bf w}^*_k(M-m+1)$, ${{\bf{w}}_k}(m)$ is the $m$-th element of ${{\bf{w}}_k}$, and $*$ denotes the conjugate operator. The outputs of transmit beamspace MIMO radar received signal can be modelled \cite{25} by substituting $[{\bf{W}},{\bf{\bar W}}]$ into \eqref{f}. Note that ${\bf \bar w}_k^H{\bm \alpha (\theta)} = {e^{ - j\frac{{2\pi }}{\lambda }(M - 1){d_t}\sin \theta }}{\left( {{{\bf{w}}_k^H}{\bm{\alpha }}(\theta )} \right)^*}$, the received signal for $l$-th target corresponding to ${\bf W}$ is the same as its received signal counterpart using ${\bf \bar W}$ up to a phase rotation of $\angle \sum\limits_{k = 1}^K {{\bf{w}}_k^H{\bm{\alpha }}({\theta _l})}  - \angle \sum\limits_{k = 1}^K {{\bf{\bar w}}_k^H{\bm{\alpha }}({\theta _l})} $, where $\angle (\cdot)$ is the phase operator. This phase difference can be applied as a look-up table finding DOA. Consequently, the RIP that enables the targets DOA estimation is enforced at the cost of complex transmit beamspace matrix design. In the following, a polynomial rooting-based method is introduced to conduct search-free DOA estimation based on the fact that ${\bf \hat X} = {\bf W}^H{\bf \hat A}$. The complex design of the transmit beamspace matrix is avoided.

First, using least-squares (LS) to compute the solution as ${\bf \hat A} = ({\bf W}^H)^{\dag}{\bf \hat X}$, where $(\cdot)^\dag$ is the pseudo-inversion. Nevertheless, the number of constraints is less than the number of variables ($K \le M$), which means that the LS solution is infinite. To tackle this problem, the condition that ${\bf A}$ is a Vandermonde matrix with ones at the first row must be utilized to narrow the solution space. Inspired by root-MUSIC method \cite{24,26,41}, the $l$-th column of ${\bf A}$ can be regarded as the powers of a complex generator $z_l$ with amplitude one. Note that ${\bf{X}}\triangleq \left[ {{{\bf{x}}_1},{{\bf{x}}_2},\cdots,{{\bf{x}}_L}} \right]$ and ${{\bf{x}}_l} = {{\bf{W}}^H}{\bm{\alpha }}({\theta _l})$, the matrix multiplication becomes the evaluation of a polynomial, given by ${{\bf{x}}_l} = {{\bf{W}}^H}{\bf p}(z_l)$, or equivalently, by
\begin{equation}
{\bf p}^H(z_l){\bf V}_l = {\bf 0}\label{qe}
\end{equation}
where ${{\cal C}({\bf p}(z_l))}\triangleq \{ {\bf{p}}(z_l) \in {\mathbb C}^{M}| {\bf{p}}(z_l) = {\left[ {1,z_l,\cdots,{z_l^{M - 1}}} \right]^T}, z_l \in {\mathbb C}\} $, ${\bf{V}}_l = {\bf{W}} - {\left[ {{{\bf{x}}_l},{{\bf{0}}_{K \times \left( {{\rm{M - 1}}} \right)}}} \right]^T}$, and ${\bf 0}$ is a vector or matrix consists of zeros. Using the definition of nullspace, i.e., ${\cal N} ({\bf V}_l^H) \triangleq \{ {\bf b} \in {\mathbb C}^{M}| {\bf V}_l^H{\bf b} = 0\}$, the null constraints in \eqref{qe} indicate that $z_l$ can be found in the intersection of ${{\cal C}({\bf p}(z_l))}$ and ${\cal N} ({\bf V}_l^H)$, or equivalently, the root of a polynomial with degree of $M-1$ and coefficients defined by the rows of ${\bf V}_l$. The compact form can be written as
\begin{equation}
{\left[ {\begin{array}{*{20}{c}}
{\rm{1}}\\
{{z_l}}\\
 \vdots \\
{z_l^{M - 1}}
\end{array}} \right]^H}\left[ {\begin{array}{*{20}{c}}
{{v_{11}}}&{{v_{12}}}& \cdots &{{v_{1K}}}\\
{{v_{21}}}&{{v_{22}}}& \cdots &{{v_{21}}}\\
 \vdots & \vdots & \ddots & \vdots \\
{{v_{M1}}}&{{v_{M2}}}& \cdots &{{v_{MK}}}
\end{array}} \right] = \left[ {\begin{array}{*{20}{c}}
0\\
0\\
 \vdots \\
0
\end{array}} \right]^T\label{poly}
\end{equation}
where $v_{mk}$ is the $(m,k)$-th element of ${\bf V}_l$. The left side describes $K$ different polynomials simultaneously. Obviously, \eqref{poly} holds if and only if
\begin{equation}
F({z_l}) \triangleq ||{\bf p}^H(z_l){\bf V}_l||^2 = 0 \label{cost}
\end{equation}
where $||\cdot||$ is the Euclidean norm. Note that $F({z_l}) = {{\bf{p}}^H}({z_l}){\bf{V}}_l{{{\bf{V}}}_l^H}{\bf{p}}({z_l})$ defines a polynomial function of degree $2(M-1)$, $z_l$ satisfies \eqref{cost} if and only if $z_l$ is the root of $F({z_l})$. The estimation of $z_l$ in \eqref{qe} is thus converted into the polynomial rooting of $F({z_l})$. Like root-MUSIC, we use the constraint $|z_l| = 1$ to select the root closest to the unit circle as $\hat z_l$, and $\theta_l$ is computed by ${\hat{\theta} _l} = \arcsin [j\lambda \ln(\hat{z_l})/2\pi {d_t}]$.

To further demonstrate the designed polynomial $F({z_l})$, we can rewrite \eqref{cost} as $\min \limits_{z_l} {{\bf{p}}^H}({z_l}){\bf{V}}_l{{{\bf{V}}}_l^H}{\bf{p}}({z_l})$ , since the Euclidean norm is by definition nonnegative. Note that ${\bf{p}}({z_l})$ and ${\bm{\alpha }}({\theta _l})$ differ only in the amplitude, the structure of ${{\bf{p}}^H}({z_l}){\bf{V}}_l{{{\bf{V}}}_l^H}{\bf{p}}({z_l})$ is identical to the structure of the transmit beampattern in MIMO radar. Hence, $F({z_l})$ describes the power transmitted into the direction $\theta _l$, given by
\begin{equation}
\min \limits_{z_l} {{\bf{p}}^H}({z_l}){\bf{V}}_l{{{\bf{V}}}_l^H}{\bf{p}}({z_l})  \Leftrightarrow  \min \limits_{\theta_l} {{\bm{\alpha }}^H}({\theta _l}){\bf{V}}_l{{{\bf{V}}}_l^H}{\bm{\alpha }}({\theta _l})\label{minpattern}
\end{equation}

Consequently, the minimization of \eqref{minpattern} points to the lowest power distribution in the beampattern generated by ${\bf V}_l$, and it is visualized as a unique deep null. This property can be explained by the idea of well-known GSC \cite{27}. Recall that ${\bf{V}}_l = {\bf{W}} - {\left[ {{{\bf{x}}_l},{{\bf{0}}_{K \times \left( {{\rm{M - 1}}} \right)}}} \right]^T}$, the sparse matrix whose first row is ${\bf x}_l$ acts as a block matrix. The cancellation therefore generates a deep null at ${\theta}_l$ and the polynomial in \eqref{cost} becomes zero when the deep null is achieved. The procedures of the proposed DOA estimation method is summarized in Table.~\ref{alg1}.

It is worth noting that the optimal value in \eqref{minpattern} is unique and achievable for any structure of the steering vector ${\bm{\alpha }}({\theta _l})$, since the objective function is surely convex. As a special case, the ULA enables the polynomial rooting method for solving \eqref{minpattern}. Essentially, the proposed search-free DOA estimation method can be generalized to scenarios with arbitrary transmit array configuration, e.g., linear or planar, uniform or non-uniform and sparse or non-sparse. Using CVX toolbox, the convex optimization problem can be solved very efficiently. Meanwhile, our proposed method can be conducted without knowing the coordinates of the receive elements ${x_n}$. The requirement of the array configuration is relaxed significantly to provide more flexibility for array design.
\begin{table}
\centering
\begin{threeparttable}
\caption{Summarization of the proposed algorithm} \label{alg1}
\centering
\begin{tabular}{p{0.2\columnwidth}p{0.8\columnwidth}}
\toprule[1.5pt]
\multicolumn{2}{l}{The DOA estimation procedures for TB MIMO Radar} \\
\hline
\textbf {Initialization} & Received signal during a single CPI ${\bf Y}_w$ from \eqref{f}\\
\textbf {Step~1} & Build the tensor ${\cal Y} =  [[ {{\bf{X}},{\bf{B}},{\bf{C}}}]]+ {\cal N}$;\\
\textbf {Step~2} & Decompose ${\cal Y}$ via \eqref{ALS} and denote the first factor matrix as ${\bf X}$;\\
\textbf {Step~3} & For $l = 1,\cdots,L$;\\
\textbf {Step~4} & Build a new $M \times K$ matrix ${\bf{V}}_l = {\bf{W}} - {\left[ {{{\bf{x}}_l},{{\bf{0}}_{K \times \left( {{\rm{M - 1}}} \right)}}} \right]^T}$;\\
\textbf {Step~5} & Construct the polynomial function $F(z_l)$ via \eqref{cost};\\
\textbf {Step~6} & Compute the root of $F(z_l)$ and select the root closest to the unit circle as $\hat z_l$;\\
\textbf {Step~7} & Estimate $\theta_l$ via ${\hat{\theta} _l} = \arcsin [j\lambda \ln(\hat{z_l})/2\pi {d_t}]$;\\
\textbf {Step~8} & Return to Step.~3 until $L$ targets DOA are estimated;\\
\bottomrule[1.5pt]
\end{tabular}
    \end{threeparttable}
\end{table}

\section{Simulation Results}
In this section, we present four simulation examples to evaluate the DOA estimation performance of the proposed method regarding root mean square error (RSME) and probability of resolution. ESPRIT\cite{9}, TENSOR\cite{8}, TB-ESPRIT\cite{11} and TB-TENSOR\cite{25} are given for comparison. Throughout the simulations, a MIMO radar with $M = 10$ and $N=10$ elements is assumed. Let ${d_t} = {\lambda}/2$, and the receive elements are randomly spaced in a linear array with aperture of $5\lambda $. $\sigma_l^2$ is chosen from a standard Gaussian distribution as a complex value. The normalized Doppler shifts are $f_1 = 0.1$ and $f_2 = -0.25$. The number of pulses in a single CPI is $Q = 64$. The number of Monte Carlo trials is 500. The orthogonal waveforms used here are ${S_m}(t) = \sqrt {\frac{1}{{{T}}}} {e^{j2\pi \frac{m}{{{T}}}t}}$.

When applying methods of ESPRIT and TENSOR, the transmit beamspace matrix is the identity matrix. The methods of TB-ESPRIT, TB-TENSOR use identical transmit beamspace matrix $[{\bf{W}},{\bf{\bar W}}]$ to enforce RIP while our proposed method applies only ${\bf{W}}$ as the transmit beamspace matrix. The emitted energy in transmit beamspace MIMO is focused on ${\Theta :[ -15^{\circ}, 15^{\circ}]}$ (see black line in Fig.~\ref{fig4}). For the first three examples, two targets at ${\theta _l} = [ -15^{\circ}, 15^{\circ}]$ are assumed. For the last example, two targets are closely spaced at ${\theta _l} = [ 10^{\circ}, 11^{\circ}]$.

\begin{figure}
\centerline{\includegraphics[width=0.75\columnwidth]{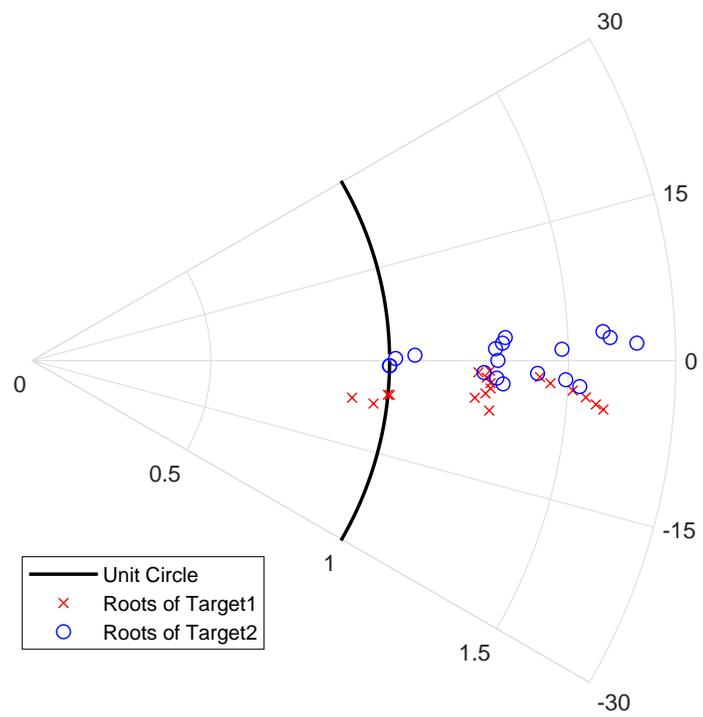}}
\caption{Polar diagram of roots of two polynomials built by \eqref{cost} for two targets at ${\theta _l} = [ -15^{\circ}, 15^{\circ}]$, SNR = 5~dB.} \label{fig3}
\end{figure}

\begin{figure}
\centerline{\includegraphics[width=1\columnwidth]{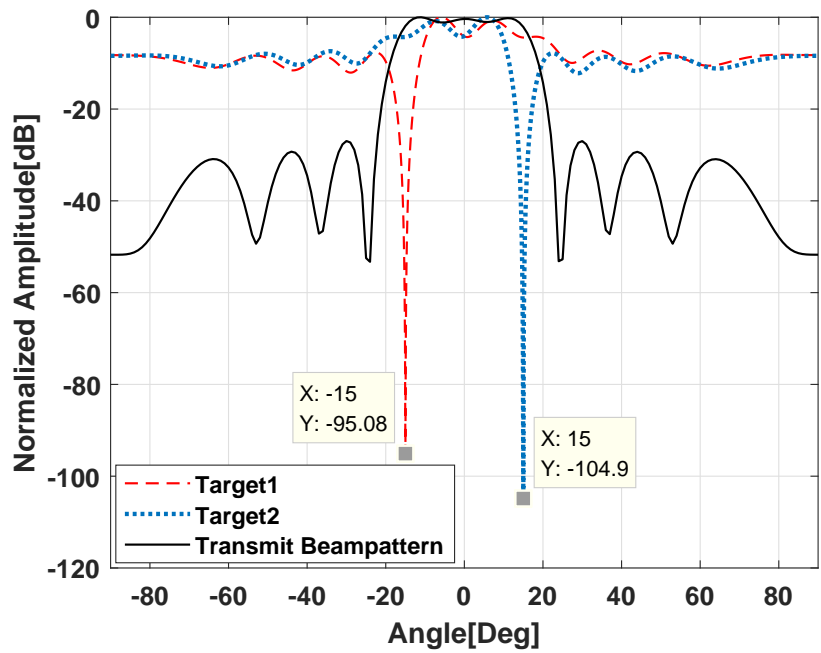}}
\caption{Transmit power distribution obtained by \eqref{minpattern} for two targets at ${\theta _l} = [ -15^{\circ}, 15^{\circ}]$, SNR = 5~dB.} \label{fig4}
\end{figure}

The polar diagram in Fig.~\ref{fig3} illustrates the roots of two different polynomials constructed by \eqref{cost} after obtaining ${\bf \hat X}$. The SNR of two targets are both 5~dB. Each polynomial corresponds to one target and the roots represent the possible solutions of target spatial angle. It can be seen from Fig.~\ref{fig3} that only one root falls on the unit circle for each target. This root can be regarded as $\hat z_l$ and the target DOA is computed. Accordingly, the transmit power distribution after GSC is performed in Fig.~\ref{fig4}. Deep nulls can be observed for both targets, whose locations reveal the target angular information, respectively. The one-to-one mapping relationship between the root of the polynomial closest to the unit circle and the target DOA is demonstrated. The effectiveness of the proposed search-free DOA estimation method is therefore verified.

\begin{figure}
\centerline{\includegraphics[width=1\columnwidth]{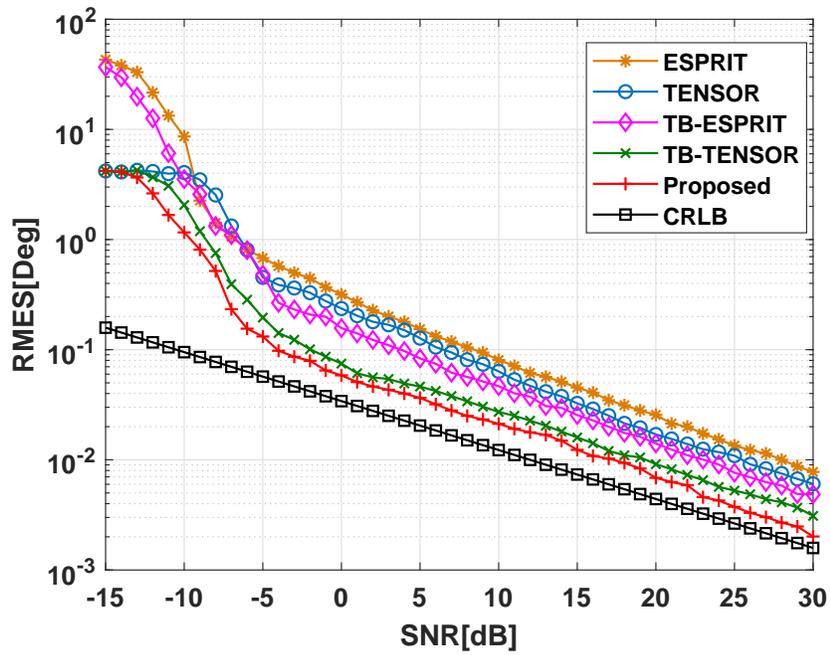}}
\caption{RMSEs of DOA estimation versus SNR, two targets at ${\theta _l} = [ -15^{\circ}, 15^{\circ}]$, 500 trials.} \label{fig1}
\end{figure}

The RMSEs of methods tested are shown in Fig.~\ref{fig1}. The RMSE decreases stably with the rise of SNR. Results of ESPRIT and TENSOR are quite similar, where TENSOR method surpasses a little. By applying the transmit beamspace technique, it can be observed that results of TB-ESPRIT and TB-TENSOR are improved as compared to their conventional MIMO counterparts, respectively. Nevertheless, the number of waveforms is doubled here to enforce RIP, which rises the complexity of transmit beamspace matrix design. In our proposed method, the design of polynomial function enables the search-free DOA estimation without additional requirement on transmit beamspace matrix. The application of the GSC demonstrates the relationship between root of polynomial nearest to the unit circle and target DOA information. The RMSE is substantially lower than those of the aforementioned methods.

\begin{figure}
\centerline{\includegraphics[width=1\columnwidth]{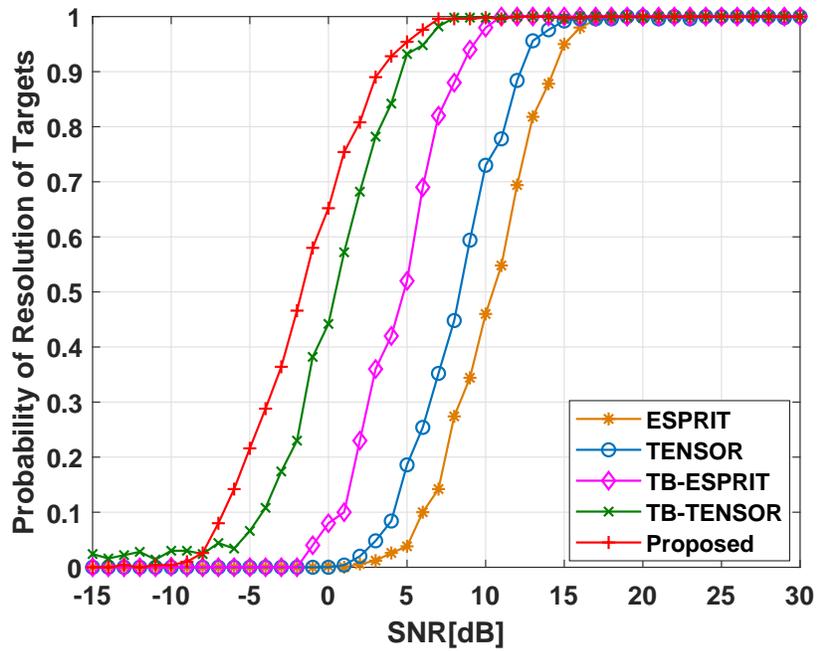}}
\caption{Probability of resolution of closely spaced targets versus SNR, two targets at ${\theta _l} = [ 10^{\circ}, 11^{\circ}]$, 500 trials.} \label{fig2}
\end{figure}

In Fig.~\ref{fig2}, the probability of resolution of two closely spaced targets is investigated. In particular, the closely spaced targets can be resolved with probability~1 when SNR is high. However, the resolution probability starts to decline with the decrease of SNR. It can be observed that our method achieves the lowest threshold. Hence, the DOA estimation performance of the proposed method surpasses other methods with better accuracy and higher resolution.

\section{Conclusion}
A search-free DOA estimation method based on tensor decomposition and polynomial rooting has been proposed to improve the accuracy and resolution for transmit beamspace MIMO radar. In the proposed method, the DOA estimation has been converted into independent polynomial rooting problems by approaching the received signal via 3-order tensor modeling and tensor decomposition. The essence of the search-free DOA estimation method is to find the deep null in the transmit beampattern after GSC. The generalization of the proposed method to arbitrary array scenario has been demonstrated. Simulation results have verified the effect performance improvement of the proposed method over conventional DOA estimation techniques for MIMO radar.

\section{Acknowledgements}
The author would like to thank Dr. Sergiy A. Vorobyov, Professor, Department of Signal Processing and Acoustics, School of Electrical Engineering, Aalto University for his constructive criticism of the manuscript. This work was supported in parts by 111 Project of China (Grant No. B14010), National Natural Science Foundation of China (Grant Nos. 61860206012, 61671065 and 31727901), and by the China Scholarship Council.

%\bibliography{mybibfile}
\bibliography{Ref}

\end{document}